\documentclass{aastex}
\usepackage{emulateapj5}
\usepackage{apjfonts}
\usepackage{epsf}

\def\calm{{\cal M}}
\def\dotm{\dot{M}_{\bullet}}
\def\ergs{\ifmmode {\rm erg~ s^{-1}} \else {\rm erg~s^{-1}}\ \fi}
\def\kms{\ifmmode {\rm km~ s^{-1}} \else {\rm km~s^{-1}}\ \fi}
\def\kabs{\kappa_{\rm abs}}
\def\mbh{M_{\bullet}}
\def\mbhc{M_{\bullet}^{\rm c}}
\def\mgii{\ifmmode Mg {\sc ii} \else Mg {\sc ii}\ \fi}
\def\oiii{[O {\sc iii}]}
\def\pp{\prime\prime}
\def\rhobh{\rho_{\bullet}}
\def\rmd{{\rm d}}
\def\rg{R_{\rm G}}
\def\siggas{\Sigma_{\rm gas}}
\def\sigsfr{\dot{\Sigma}_{\rm SFR}}
\def\sunm{M_{\odot}}

\def\lax{{$\mathrel{\hbox{\rlap{\hbox{\lower4pt\hbox{$\sim$}}}\hbox{$<$}}}$}}
\def\gax{{$\mathrel{\hbox{\rlap{\hbox{\lower4pt\hbox{$\sim$}}}\hbox{$>$}}}$}}

\journalinfo{The Astrophysical Journal Letters: in Press}

\begin{document}

\title{Suppressed star formation in circumnuclear regions in Seyfert galaxies}

\author{Jian-Min Wang, Yan-Mei Chen, Chang-Shuo Yan, Chen Hu and Wei-Hao Bian}

\affil{Key Laboratory for Particle Astrophysics, Institute of High Energy Physics, CAS,
19B Yuquan Road, Beijing 100049, China}

\slugcomment{Received 2007 January 17; accepted 2007 March 30}
\shorttitle{Suppressed Starburst in Seyfert Galaxies}
\shortauthors{Wang et al.}

\begin{abstract}
Feedback from black hole activity is widely believed to play a key role in regulating star formation 
and black hole growth. A long-standing issue is the relation between the star formation and fueling 
the supermassive black holes in active galactic nuclei (AGNs). We compile a sample of 57 Seyfert 
galaxies to tackle this issue. We estimate the surface densities of gas and star formation rates in 
circumnuclear regions (CNRs). Comparing with the well-known Kennicutt-Schmidt (K-S) law,  we find 
that the star formation rates in CNRs of most Seyfert galaxies are suppressed in this sample. Feedback 
is suggested to explain the suppressed star formation rates. 
\end{abstract}

\section{Introduction}
The implications of the well-known relations between black hole masses and bulge magnitudes (Magorrian et al. 1998), 
and the velocity dispersions (Gebhardt et al. 2000; Frarreasse \& Merrit 2000) show a coevolution of the black holes 
and their host galaxies. However, how do black holes know the evolution stage of the galaxies and how to control the 
growth of the black holes are currently understood via the the feedback from the black hole (Silk \& Rees 1998; 
Croton et al. 2006; Schawinski et al. 2006). 
Numerical simulations show two roles of feedback from the black hole activity: (1) modulating the star formation rates; 
(2) heating the medium and finally quenching the black hole activity (Di Matteo et al. 2005). The direct evidence for 
the presence of the feedback from active black holes has to be shown from observations, yet.

The main goal of the present paper is to show one piece of evidence for the feedback role in active galaxies. 
We show the AGN feedback domain, where starburst should be suppressed. We find the star formation rates
in Seyfert galaxies is significantly lower than the rates predicted by the Kennicut-Schmidt's law.
We use the cosmological parameters $H_0=75{\rm km~s^{-1}~Mpc^{-1}}$, 
$\Omega_{\rm M}=0.3$ and $\Omega_{\Lambda}=0.7$ throughout calculations.

\section{AGN feedback domain}
When the CNR medium is optically thick, namely, 
the optical depth $\tau=\kabs\siggas\ge 1$, where $\kabs$ is opacity and $\siggas$ the gas surface density, the 
radiation from the black hole activity will continuously heat the medium and blow the gas away so as to lower
the star formation rates. The condition of $\tau=1$ yields a critical density 
\begin{equation}
\siggas^{c_1}=9.0\times 10^2 \left(\kabs/5\right)^{-1}~\sunm~{\rm pc^{-2}},
\end{equation}
$\kabs$ has a mean value of 5 for the CNR medium (Semenov et al. 2003). This is feedback driven by AGN radiation. 
We note outflows from Seyfert active nucleus
have much low kinetic luminosities, typically $\sim 10^{-(3-6)}L_{\rm Bol}$ based on X-ray warmer 
absorbers (Blustin et al. 2005), where $L_{\rm Bol}$ is
the bolometric luminosity. Feedback from outflows could be thus neglected. When $\siggas>\siggas^{c_1}$, 
the AGN radiation-driven feedback will suppress the star formation.
On the other hand, AGN feedback reaches its maximum when an AGN radiates at the Eddington limit
$L_{\rm AGN}=L_{\rm Edd}=1.3\times 10^{38}(\mbh/\sunm)$erg/s.
In the case of $L_{\rm Edd}\le L_{\rm SFR}^{\rm IR}$, AGN have inefficient feedback to star formation.
With the help of ${\rm SFR}=4.5\left(L_{\rm IR}/10^{44}\ergs\right)\sunm{\rm yr}^{-1}$, 
$\siggas^{c_2}$ is given by using the
K-S law $\sigsfr=A\siggas^{\gamma}$ (Kennicutt 1998a),
\begin{equation}
\siggas^{\rm c_2}= 8.2\times 10^5M_9^{0.7}R_{200}^{-1.4}~\sunm~{\rm pc^{-2}},
\end{equation}
where $\sigsfr={\rm SFR}/\pi R^2$ is the surface density of the star formation rate, 
$A=2.5\times 10^{-4}$, $\gamma=1.4$, $M_9=\mbh/10^9\sunm$ is the black hole mass and
$R_{200}=R/200{\rm pc}$ the size of the circumnuclear star forming region. When $\siggas\ge\siggas^{\rm c_2}$,
the gas is so dense that the luminosity from star formation dominates over the AGN. We call 
\begin{equation}
\siggas^{c_1}\le \siggas\le \siggas^{c_2},
\end{equation}
the AGN feedback domain as shown in Fig. 1, in which the K-S law is broken.

The strong radiation pressure from the black hole accretion disk at Eddington limit is 
$P_{\rm AGN}\approx 1.0\times 10^{-7}M_{8}R_{200}^{-2}$dyn~cm$^{-2}$, where $M_{8}=\mbh/10^8\sunm$. The pressure 
from supernovae explosion is 
$P_{\rm SN}\approx \epsilon \sigsfr c=2.0\times 10^{-8}\dot{\Sigma}_{\rm SFR,2}\epsilon_{-3}$dyn~cm$^{-2}$, 
where $\dot{\Sigma}_{\rm SFR,2}=\sigsfr/10^2\sunm{\rm yr^{-1}kpc^{-2}}$ and $\epsilon_{-3}=\epsilon/10^{-3}$ is the 
efficiency converting the mass into radiation (Thompson et al 2005). We find $P_{\rm AGN}\ge 5P_{\rm SN}$ within  
CNRs of radius $\sim 200$pc for typical values of the parameters of $\epsilon$, $\mbh$ and $\sigsfr$. This indicates
that the radiation from AGN dominates over the local feedback from supernovae explosion. After an AGN switches on,
the star formation is suppressed and then feedback from supernovae is further weakened. The timescale of the AGN 
feedback to the starburst regions can be estimated by 
$t_{\rm FB}\sim E_{\rm gas}/f_{\rm FB}{\cal C}L_{\rm AGN}$, where $L_{\rm AGN}$ is AGN luminosity,
${\cal C}=\Delta \Omega/4\pi$ is the covering factor, 
the thermal energy $E_{\rm gas}\approx kTM_{\rm gas}/m_p$, $k$ is the Boltzmann constant, $m_p$ is the proton mass, $T$ 
is the gas temperature, $M_{\rm gas}=\pi R^2\siggas$ is the gas mass and $f_{\rm FB}$ is the feedback efficiency. We 
have 
\begin{equation}
t_{\rm FB}\sim 2.6\times 10^5f_{\rm FB,-2}^{-1}R_{200}^2T_3\Sigma_{\rm gas,4}{\cal C}_{0.5}^{-1}L_{43}^{-1}~~~{\rm yr}, 
\end{equation}
where $\Sigma_{\rm gas,4}=\Sigma_{\rm gas}/10^4{\sunm~\rm pc^{-2}}$,
$T_3=T/2\times 10^3$K, $f_{\rm FB,-2}=f_{\rm FB}/10^{-2}$, $L_{43}=L_{\rm AGN}/10^{43}\ergs$ and 
${\cal C}_{0.5}={\cal C}/0.5$ is the covering factor of the CNRs. Such a short timescale indicates that
the AGN feedback is very efficient. This is supported by a large fraction of the
post-starburst AGNs in a large Sloan Digital Sky Survey sample (Kauffmann et al. 2003). The physics behind K-S law is 
not sufficiently understood (Thompson et al. 2005; Krumholz \& McKee 2005). It is beyond the scope of the
present paper to give a quantitative description of the suppressed star formation rates. Comparing 
Seyfert galaxies with the K-S law, a universal rule of cosmic star formation, 
we may get the undergoing physics in the CNRs.

\begin{table*}[t]
\begin{center}
\centerline{\sc Table 1 The Seyfert Galaxy Sample}
\vskip 0.2cm
{\footnotesize
\begin{tabular}{lcccllccrlll}\hline
\multicolumn{12}{c}{Seyfert 1}\\ \hline

                    Object
                  & Redshift
                  & FWHM
                  & $\log\lambda L_{\lambda}$
		  & Ref.
                  & $\log \mbh$
                  & $\dot{M}_{\bullet}$
                  & $\log\siggas$
                  & $S_{\rm PAH}$
                  & $R$
                  &\multicolumn{2}{c} {$\log\sigsfr$} \\ \cline{11-12}
(1)&(2)&(3)&(4)&(5)&(6)&(7)&(8)&(9)&(10)&(11)&(12)\\ \hline
3C120		&0.033	&...	&44.17	& 2	&7.74$^{a}$	&0.24 &4.18 &  76	&0.48	&0.85	&1.32\\
IC4329		&0.016	&...	&43.32	& 2	&6.99$^{a}$	&0.03 &3.70 & 220	&0.24	&1.31	&1.35\\
MCG-2-33-34	&0.014	&1565	&42.61	& 22, 5	&6.11	&0.01         &3.13 &  54	&0.21	&0.70	&1.21\\
MCG-5-13-17	&0.013	&4000	&43.44	& 20, 1	&7.50	&0.04         &3.92 &  67	&0.19	&0.79	&1.24\\
\hline
\end{tabular}
}
\vskip 0.1cm
\parbox{5.55in}
{\baselineskip 4pt
\indent \footnotesize
NOTE.$-$Table 1 is published in its entirety in the electronic edition. A portion is shown here for guidance regarding 
its form and content.\\
\vskip 0.2cm
$^{a}$the blackhole mass are directly taken from Peterson et al. (2004).\\
$^{b}$refers to [O {\sc iii}] FWHM.\\
$^{c}$based on $\mbh-M_{\rm bulge}$ relation, F01475-0740: $M_{\rm bulge}=-18.80$; NGC 3660: $M_{\rm bulge}=-18.38$.
\vskip 0.2cm
Note.-(1) source name; (2) redshift; (3) FWHM of H$\beta$ for Seyfert 1s or stellar velocity dispersion $\sigma$ for 
Seyfert 2s (in km s$^{-1}$); (4) luminosity of 5100\AA~ deduced from extrapolation of $F_{\nu}\propto \nu^{-0.5}$
or \oiii$\lambda 5007$\AA~ (in erg~s$^{-1}$);
(5) references for columns (3) and (4) are given below, respectively; (6) black hole mass (in $\sunm$); (7) accretion 
rate (in $\sunm$yr$^{-1}$);
(8) gas surface density (in $\sunm$pc$^{-2}$); (9) surface brightness of the 3.3 $\mu$m PAH emission feature
(in unit of $10^{39}$ergs s$^{-1}$kpc$^{-2}$); (10) the scale of the starburst regions (in kpc); (11) and (12) are
the lower ($\sigsfr^{\rm L}$) and upper ($\sigsfr^{\rm U}$) limits of surface density of 
star formation rates, respectively (in $\sunm$yr$^{-1}$kpc$^{-2}$).
\vskip 0.2cm
Reference.-(1) NED; (2) Peterson et al. (2004); (3) Spinogilio et al. (1995); (4) Doroshenko \& 
Terebezh (1979); (5) Kinney et al. (1993); (6) Nelson \& Whittle (1995); (7) Dahari \& Robertis (1988); (8) Lipari 
et al. (1991); (9) Corral et al. (2005); (10) Kirhakos \& Steiner (1990); (11) Visvanathan \& Griersmith (1977); 
(12) Cid Fernandes et al. (2004); (13) Gu \& Huang (2002);
(14) Kailey \& Lebofsky (1988); (15) Heraudeau \& Simien (1998); (16) Bassani (1999); (17) Whittle et al. (1988);
(18) Whittle (1992); (19) Garcia-Rissmann et al. (2005); (20) Crenshaw et al. (2003); (21) Marzini et al. (2003); 
(22) Veron-Cetty et al. (2001); (23) Postman \& Lauer (1995).
}
\end{center}
\end{table*}

We have to point out here that the short feedback time does NOT mean the same timescale of the starburst. The
present $t_{\rm FB}$ means the starburst rates will be suppressed once AGN is triggered and make it possible
for AGN and starburst coexist.

\section{Appearance of feedback in Seyfert galaxies}
For the goal to test the above scenario, we compile 57 Seyfert 
galaxies (Imanishi 2002; Imanishi 2003; Imanishi \& Wada 2004). The star formation rates in CNRs of Seyfert galaxies 
can be traced by several indicators, particularly, PAH features at 3.3, 6.2, 7.7, 8.6 and 11.2$\mu$m,
which radiate from vibration of PAH grains containing about 50 carbon atoms. Among the features, 3.3$\mu$m 
emission is intrinsically strong and less affected by broad silicate dust absorption (Imanishi 2002). We choose 
3.3$\mu$m emission as an indicator of the star formation rate. We convert the PAH emission into IR luminosity via 
$L_{\rm IR}=10^3L_{\rm PAH}$ relation with a scatter by a factor of 2-3 for pure star formation (Imanishi 2002). 
Since some PAH grains would be destroyed by EUV and X-ray photons from the central engine, we have the lower limit 
of the surface density of the star formation rates 
\begin{equation}
\sigsfr^{\rm L}= 35.8L_{\rm PAH,41}R_{200}^{-2}~({\sunm{\rm yr^{-1}kpc^{-2}}}),
\end{equation}
by using the relation of the star formation rate and the infrared luminosity (eq. 7) (Kennicutt 1998a), where 
$L_{\rm PAH,41}=L_{\rm PAH}/10^{41}\ergs$. On the other hand, the infrared emission from Seyfert galaxies
covers the contribution from starburst and reprocessing radiation from AGNs, we have the upper limit of the surface 
density of the star formation rates
\begin{equation}
\sigsfr^{\rm U}= 35.8L_{\rm FIR,44}R_{200}^{-2}~({\sunm{\rm yr^{-1}kpc^{-2}}}),
\end{equation}
where $L_{\rm FIR,44}=L_{\rm FIR}/10^{44}\ergs$ is the observed far-IR luminosity. 
We take the geometric average $\sigsfr=\left(\sigsfr^{\rm L}\sigsfr^{\rm U}\right)^{1/2}$ 
and the error bars correspond to $\sigsfr^{\rm L}$ and $\sigsfr^{\rm U}$. We have to stress this average
only represents the central value of logarithm of $\sigsfr^{\rm U}$ and $\sigsfr^{\rm L}$ and the upper and lower limits
of $\sigsfr$ are the most important. Table 1 gives the sample of Seyfert 
galaxies, which have been observed through IRTF SpeX or Subaru IRCS with spatial 
resolution of $0.9^{\pp}-1.6^{\pp}$. 

\begin{figure*}[t]
\centerline{\includegraphics[angle=-90,width=11.0cm]{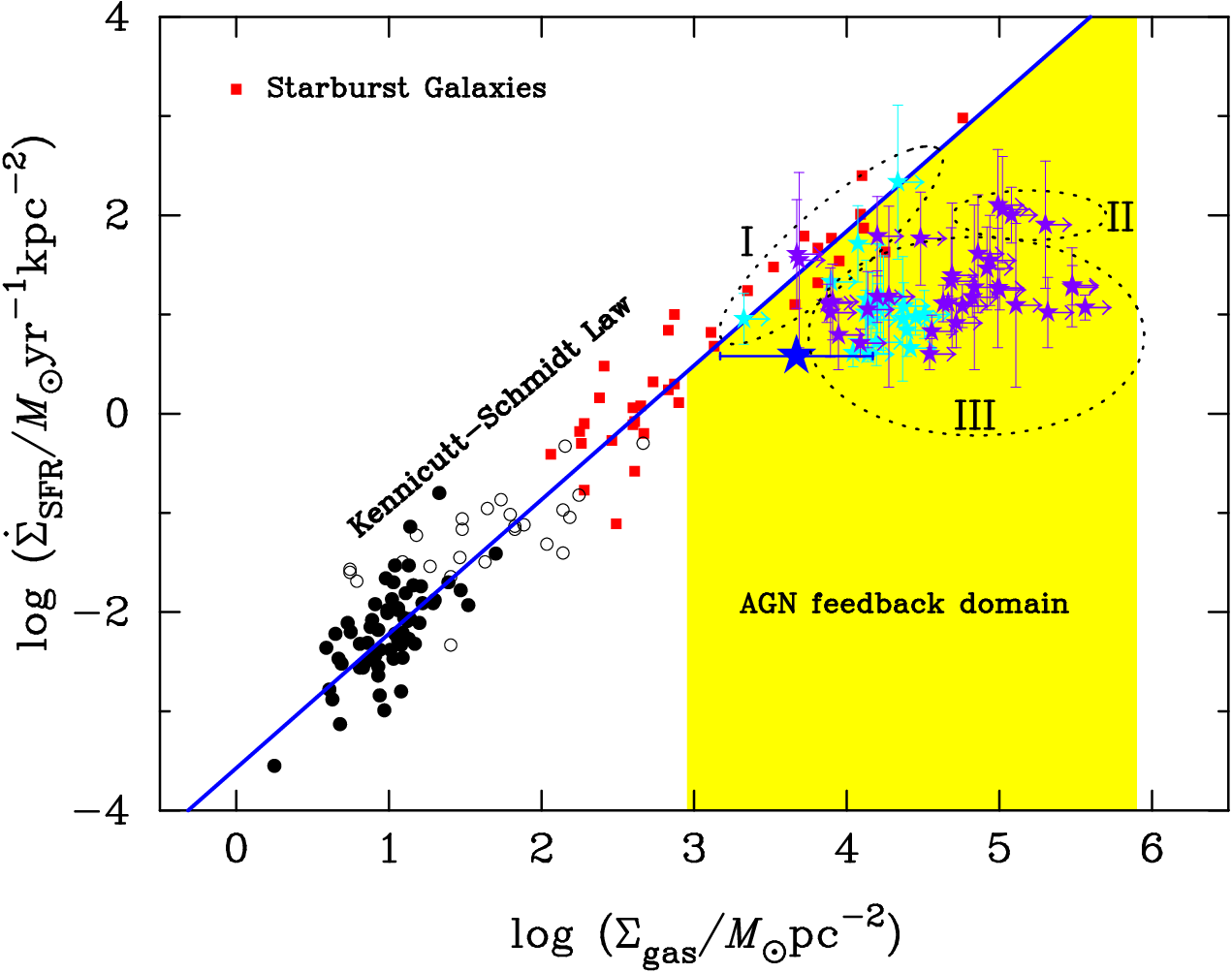}} 
\caption{The plot of gas and star formation rate surface densities.
The yellow region is the AGN feedback domain given by $\siggas^{c_1}\le \siggas\le \siggas^{c_2}$. 
The Compton thick region has $\siggas\ge 8.0\times 10^3\sunm~{\rm pc^{-2}}$ (i.e. $N_{\rm H}\ge 10^{24}{\rm cm^{-2}})$.
The red squares are starburst galaxies taken from Kennicutt (1998b).  The cyan and blue-magenta stars 
are Seyfert 1 and 2 galaxies, respectively. The blue star is NGC 3227, in which the star formation rate
is 0.05$\sunm~{\rm yr}^{-1}$ and the gas mass $M_{\rm gas}=(2-20)\times 10^8\sunm$ within 65pc taken from 
Davies et al. (2006). 
}
\label{fig1}
\end{figure*}

For Seyfert 1 galaxies, we estimate  
$L_{\rm Bol}=9L_{5100}$, where $L_{5100}$ is the luminosity at 5100\AA~ and then the accretion rate
$\dot{\mbh}=L_{\rm Bol}/\eta c^2$, where $\eta=0.1$ is the accretion efficiency. The black hole masses are estimated 
from the empirical relation of reverberation mapping (Kaspi et al. 2000), or directly taken from the mapping
(Peterson et al. 2004). We assume that the potential of the total mass
within the CNRs controls the massive disk fueling to the black hole, where star formations are taking place either.
Assuming the Keplerian rotation, the surface density of the disk is 
\begin{equation}
\Sigma_{\rm tot}=2.1\times 10^6 \alpha_{0.1}^{-4/5}\dot{M}_{\bullet,1}^{3/5}f_{\bullet}^{-1/5}M_{8}^{1/5}R_{200}^{-3/5}~~~~
\sunm~{\rm pc^{-2}},
\end{equation}
given by the disk model (King et al. 2002; Yi \& Blackman 1994; Tan 2005), where the opacity  $\kabs=5$ in 
this region, $f_{\bullet}$ is the ratio of the black hole 
mass to the total, $\dot{M}_{\bullet,1}=\dot{M}_{\bullet}/1.0\sunm{\rm yr^{-1}}$ and $\alpha_{0.1}=\alpha/0.1$ is the 
viscosity (Shakura \& Sunyaev 1973). This estimation is the lower limit since we replace the infalling mass rates in 
CNRs by black hole accretion rates. The gas surface density of the disk $\siggas=f_{\rm g}\Sigma_{\rm tot}$
\begin{equation}
\siggas=1.0\times 10^5 f_{\rm g,0.05}\alpha_{0.1}^{-4/5}
\dot{M}_{\bullet,1}^{3/5}f_{\bullet}^{-1/5}M_{8}^{1/5}R_{200}^{-3/5}~~~~\sunm~{\rm pc^{-2}},
\end{equation}
where  $f_{\rm g,0.05}=f_{\rm g}/0.05$ is the gas fraction to the total. Considering the disk is located inside
the bulge, we have $f_{\rm g}=M_{\rm gas}/M_{\rm disk}> M_{\rm gas}/M_{\rm Bulge}
=\left(M_{\rm gas}/M_{\rm dust}\right)\left(M_{\rm dust}/\mbh\right)\left(\mbh/M_{\rm Bulge}\right)$,
where $M_{\rm disk}$ is the total mass of the disk, $M_{\rm gas}/M_{\rm dust}$ is the gas-to-dust mass
ratio and $M_{\rm Bulge}\approx 10^3\mbh$ is the bulge mass (Kormendy \& Gebhardt 2001; McLure \& Dunlop 2002).
It has been found that the dust mass in PG quasars is comparable with in Seyfert galaxies (Spinoglio et al. 2002; 
Haas et al. 2003). The mean value of gas-to-dust mass ratio is $\langle M_{\rm gas}/M_{\rm dust}\rangle\sim 250$ 
(Haas et al. 2003). We estimated dust mass from 
$M_{\rm dust}\sim 4.78f_{\rm 100\mu}D_{\rm L}^2\left[\exp(143.38/T_{\rm dust})-1\right]\sunm$, and the dust 
temperature is estimated by $T_{\rm dust}=(1+z)\left[0.5-82/\ln(0.3f_{60\mu}/f_{100\mu})\right]$K, where
$f_{100\mu}$ and $f_{60\mu}$ are the fluxes at 100$\mu m$ and $60\mu m$ in unit of Jy, respectively, $D_{\rm L}$ 
is the luminosity distance in unit of Mpc (Evans et al. 2005). We find the mean value of  
$\langle M_{\rm dust}/\mbh\rangle=0.2\pm 0.2$ in our sample. So we have $f_{\rm g}\ge 0.05$ as a lower limit in this 
paper. Thompson et al. (2005) used $f_{\rm g}=0.1$.
We note $\siggas\propto f_{\bullet}^{-1/5}$, resulting in uncertainties of $\siggas$ by a factor of 4 
for $f_{\bullet}=10^{-3}-1$. $\alpha=0.1$ is used for all Seyfert galaxies.

For Seyfert 2 galaxies, dusty tori obscure the active regions. We estimate the bolometric 
luminosity from $L_{\rm Bol}=3500L_{\rm [O~III]}$ with a mean uncertainty of 0.38 dex (Heckman et al. 2004), 
where $L_{\rm [O~III]}$ is the \oiii${\lambda5007}$ luminosity, and hence 
the black hole accretion rates. The black hole masses are estimated through the $\mbh-\sigma$ 
relation (Tremaine et al. 2002), where the dispersion velocity $\sigma={\rm FWHM([O}~${\sc iii}$])/2.35$ if the dispersion
velocity is not available.

Fig. 1 shows the $\siggas-\sigsfr$ plot of Seyfert CNRs. We find that CNR gas surface densities of Seyfert galaxies 
are located within the AGN feedback domain. There are clearly three branches in the 
figure, separating the Seyfert galaxies, when $\siggas^{c_2}>\siggas>\siggas^{c_1}$. Seyfert galaxies 
marked in Zone I still satisfy the K-S law. Those (Mrk 273, Mrk 938, NGC 5135 and NGC 1068) marked in Zone II are 
located between the K-S law and Zone III. These are ultra-luminous infrared galaxies, or mixed with strong starbursts. 
The main energy sources in the CNRs are in a transition state from a starburst to an AGN in these galaxies. The fraction
of the transiting galaxies is only $4/57\sim 1/10$. Though the completeness of the present sample is uncertain,
this fraction implies that the transition is quite short and indicated by the feedback timescale from equation (4). 
The Seyfert galaxies in Zone III are undergoing suppressed star formation strongly, 
being 1-2 orders lower than that predicted by the K-S law.  The suppressed $\sigsfr$ is obviously caused by the 
feedback. Galaxies obeying the K-S law are powered by nuclear energy from stars, however gravitational energy 
released from accretion onto the black holes is powering AGNs if a transition from starburts to active galaxies happens. 
With the dissipation of CNR gas due to star formation and accretion onto the black holes, $\siggas$ is decreasing and
the galaxies may return to the K-S law once AGNs switch off. Such a behavior likes evolution of stellar energy sources 
in the Hertzprung-Russell diagram.

It has been found that black hole duty cycles follow the history of star formation rate density (Wang et al. 2006). The 
above scenario then implies that both the black hole activities and starbursts are episodic (Davies et al. 2006). The 
multiple cycles of the black holes and starbursts  make it impossible to measure the time delay between the two episodes. 
However the stellar synthesis may tell the star formation history and then give the black hole activity history.

\section{Conclusions and Discussions}
We find direct evidence for the feedback from active black holes in Seyfert galaxies. Once a black hole is triggered,
the feedback will significantly suppress the starbursts within a quite short timescale of a few $10^5$years. The duty 
cycles of Seyfert galaxies strongly indicate there is an efficient way to frequently trigger black holes and quench 
starbursts.

The data presented in this paper are only lower limits of the gas densities. Future VLT (Very Large Telescope) and ALMA 
(Atacama Large Millimiter Array) measurements of the star formation rates 
and gas densities will finally identify roles of the feedback from the black hole activities.

\acknowledgements{The helpful comments from the referee are acknowledged.
The authors are grateful to R. C. Kennicutt, L. C. Ho, S. N. Zhang and X.-Y. Xia for useful discussions.
We appreciate the stimulating discussions among the members of IHEP (Institute of High Energy Physics) AGN group.
J.-M.W. thanks the Natural Science Foundation of China for support via NSFC-10325313 and 10521001, CAS key project via 
KJCX2-YW-T03.}

\def\calm{{\cal M}}
\def\dotm{\dot{M}_{\bullet}}
\def\ergs{\ifmmode {\rm erg~ s^{-1}} \else {\rm erg~s^{-1}}\ \fi}
\def\kms{\ifmmode {\rm km~ s^{-1}} \else {\rm km~s^{-1}}\ \fi}
\def\kabs{\kappa_{\rm abs}}
\def\mbh{M_{\bullet}}
\def\mbhc{M_{\bullet}^{\rm c}}
\def\mgii{\ifmmode Mg {\sc ii} \else Mg {\sc ii}\ \fi}
\def\oiii{[O {\sc iii}]}
\def\pp{\prime\prime}
\def\rhobh{\rho_{\bullet}}
\def\rmd{{\rm d}}
\def\rg{R_{\rm G}}
\def\siggas{\Sigma_{\rm gas}}
\def\sigsfr{\dot{\Sigma}_{\rm SFR}}
\def\sunm{M_{\odot}}

\def\lax{{$\mathrel{\hbox{\rlap{\hbox{\lower4pt\hbox{$\sim$}}}\hbox{$<$}}}$}}
\def\gax{{$\mathrel{\hbox{\rlap{\hbox{\lower4pt\hbox{$\sim$}}}\hbox{$>$}}}$}}

\vskip 0.5cm

{\footnotesize

\begin{center}
\begin{tabular}{lcccllccrlll}
\multicolumn{12}{c}{{\small Table 1 The Seyfert Galaxy Sample}}\\ 
\multicolumn{12}{c}{~}\\ \hline
\multicolumn{12}{c}{Seyfert 1}\\ \hline

                    Object
                  & Redshift
                  & FWHM
                  & $\log\lambda L_{\lambda}$
		  & Ref.
                  & $\log \mbh$
                  & $\dot{M}_{\bullet}$
                  & $\log\siggas$
                  & $S_{\rm PAH}$
                  & $R$
                  &\multicolumn{2}{c} {$\log\sigsfr$} \\ \cline{11-12}
(1)&(2)&(3)&(4)&(5)&(6)&(7)&(8)&(9)&(10)&(11)&(12)\\ \hline
3C120		&0.033	&...	&44.17	& 2	&7.74$^{a}$	&0.24 &4.18 &  76	&0.48	&0.85	&1.32\\
IC4329		&0.016	&...	&43.32	& 2	&6.99$^{a}$	&0.03 &3.70 & 220	&0.24	&1.31	&1.35\\
MCG-2-33-34	&0.014	&1565	&42.61	& 22, 5	&6.11	&0.01         &3.13 &  54	&0.21	&0.70	&1.21\\
MCG-5-13-17	&0.013	&4000	&43.44	& 20, 1	&7.50	&0.04         &3.92 &  67	&0.19	&0.79	&1.24\\
Mrk79		&0.022	&...	&43.72	& 2	&7.72$^{a}$	&0.08 &4.00 &  71	&0.32	&0.82	&1.31\\
Mrk335		&0.025	&...	&43.86	& 2	&7.15$^{a}$	&0.14 &3.94 &  36	&0.37	&0.52	&0.68\\
Mrk509		&0.035	&...	&44.28	& 2	&8.15$^{a}$	&0.30 &4.31 &  62	&0.51	&0.76	&1.24\\
Mrk530  	&0.029	&6560	&44.04	& 20, 1	&8.35	&0.17         &4.25 &  67	&0.42	&0.79	&1.15\\
Mrk618		&0.035	&3018	&44.00	& 21, 4	&7.65	&0.16         &4.04 &  44	&0.51	&0.61	&1.59\\
Mrk704		&0.030	&5684	&43.53	& 21, 1	&7.88	&0.05         &3.84 &  32	&0.44	&0.47	&0.76\\
Mrk817		&0.031	&...	&43.82	& 2	&7.69$^{a}$	&0.11 &3.97 &  78	&0.45	&0.86	&1.42\\
Mrk1239		&0.019	&1075	&43.84	& 22, 3	&6.65	&0.11         &4.02 &  47	&0.17	&0.64	&1.70\\
NGC863		&0.027	&...	&43.81	& 2	&7.68$^{a}$	&0.10 &4.00 &  33	&0.40	&0.48	&0.95\\
NGC931		&0.016	&1830	&43.70	& 20, 1	&7.01	&0.08         &3.93 &  61	&0.24	&0.75	&1.55\\
NGC2639		&0.011	&3100	&43.77	& 20, 1	&7.51	&0.09         &4.17 &  23	&0.16	&0.33	&1.58\\
NGC4235		&0.008	&7600	&43.51	& 20,11	&8.11	&0.05         &4.22 &  49	&0.12	&0.66	&0.66\\
NGC4253		&0.013	&1630	&43.41	& 20, 4	&6.70	&0.04         &3.87 & 230	&0.12	&1.33	&2.10\\
NGC5548		&0.017	&...	&43.51	& 2	&7.83$^{a}$	&0.05 &3.96 &  61	&0.26	&0.75	&1.14\\
NGC5940		&0.034	&5240	&44.07	& 20, 1	&8.18	&0.19         &4.20 &  41	&0.49	&0.58	&1.10\\
NGC7469		&0.016	&...	&43.72	& 2	&7.08$^{a}$	&0.08 &4.14 & 390	&0.12	&1.56	&3.11\\
\hline
\multicolumn{11}{c}{Seyfert 2}\\ \hline
                  Object
                  & Redshift
                  & $\sigma$
                  & $\log L_{\rm [O~ III]}$
                  & Ref.
                  & $\log\mbh$
                  & $\dot{M}_{\bullet}$
                  & $\log\siggas$
                  & $S_{\rm PAH}$
                  & $R$
                  &\multicolumn{2}{c} {$\log\sigsfr$} \\ \cline{11-12}
(1)&(2)&(3)&(4)&(5)&(6)&(7)&(8)&(9)&(10)&(11)&(12)\\ \hline
F01475-0740	&0.017	&...	        &41.69	&13	&7.55$^{c}$	&0.30 &4.36&  50	&0.26	&0.66	&0.99\\
F04385-0828	&0.015	&907$^{b}$	&40.12	& 7,13	&8.77	&0.01         &3.71&  60	&0.22	&0.74	&1.51 \\
F15480-0344	&0.030	&664$^{b}$	&42.95	& 8,13	&8.22	&5.46         &5.12&  50	&0.43	&0.66	&1.37  \\
IC3639		&0.011	& 95	        &42.11	&19,13	&6.83	&0.80         &4.66& 113	&0.13	&1.02	&2.21\\
MCG-3-34-64	&0.017	&155	        &42.32	&12,13	&7.69	&1.30         &4.79&  50	&0.24	&0.66	&1.87\\
Mrk34		&0.015	&570$^{b}$	&43.04	& 7, 7	&7.96	&6.78         &5.36&  95	&0.17	&0.94	&1.20 \\
Mrk78		&0.037	&172	        &42.22	& 6,17	&7.87	&1.03         &4.63& 102	&0.41	&0.97	&1.37\\ 
Mrk273		&0.038	&211	        &42.39	&19,16	&8.22	&1.52	&4.79& 377	&0.42	&1.54	&2.66\\ 
Mrk334		&0.022	&250$^{b}$	&41.29	& 7,13	&6.52	&0.12	&4.00& 265	&0.19	&1.39	&2.19  \\
Mrk463		&0.051	&545$^{b}$	&43.44	&18,18	&7.88	&16.82	&5.28&  81	&0.56	&0.88	&1.67  \\
Mrk477		&0.038	&370$^{b}$	&43.54	&18,13	&7.20	&21.57	&5.28& 135	&0.42	&1.10	&1.49  \\
Mrk573		&0.017	&123	        &42.00	& 6,13	&7.28	&0.62	&4.52&  50	&0.24	&0.66	&1.16\\
Mrk938		&0.019	&330$^{b}$	&42.77	& 7,13	&7.00	&3.66	&4.88& 570	&0.28	&1.72	&2.28 \\
Mrk993          &0.015  &392$^{b}$      &40.87  & 9,7   &7.30   &0.05  &3.86&  30     &0.22   &0.44   &0.82\\
NGC262		&0.015	&118	        &41.91	& 6,13	&7.21	&0.51	&4.47& 120	&0.22	&1.04	&1.19\\
NGC513		&0.020	&152	        &40.60	& 6,13	&7.65	&0.02	&3.69&  40	&0.30	&0.57	&1.47\\
NGC1068		&0.004	&151	        &42.65	& 6,13	&7.64	&2.73	&5.10& 198	&0.15	&1.26	&2.55\\
NGC1194		&0.013	&396$^{b}$	&40.84	&10,14	&7.32	&0.04	&3.89&  40	&0.19	&0.57	&0.85 \\
NGC1241		&0.014	&136	        &42.47	&19,13	&7.46	&1.83	&4.91&  20	&0.18	&0.27	&1.92\\
NGC1320		&0.010	&116	        &40.96	& 6,13	&7.18	&0.06	&4.00&  90	&0.15	&0.92	&1.44\\
NGC1667		&0.015	&173	        &41.91	& 6,13	&7.88	&0.51	&4.64&  30	&0.19	&0.44	&2.10\\
NGC2992		&0.008	&158	        &41.92	& 6,16	&7.72	&0.52	&4.74& 133	&0.12	&1.09	&2.00\\
NGC3660		&0.012	&...	        &40.91	&13	&7.33$^{c}$&0.05&3.94&  50	&0.19	&0.66	&1.43\\
NGC3786		&0.009	&142	        &41.52	& 6, 7	&7.53	&0.20	&4.43&  91	&0.13	&0.92	&1.30\\ 
NGC4388		&0.008	&119	        &41.68	& 6,13	&7.22	&0.30	&4.49&  50	&0.12	&0.66	&2.12\\
NGC4501		&0.008	&171	        &39.80	&15,13	&7.86	&0.01	&3.49&  50	&0.12	&0.66	&2.43\\
NGC4968		&0.010	&105	        &42.37	&19,16	&7.01	&1.44	&4.79& 120	&0.15	&1.04	&1.45\\
NGC5135		&0.014	&128	        &42.31	&19,13	&7.35	&1.26	&4.82& 360	&0.16	&1.52	&2.59\\
NGC5252		&0.023	&190	        &41.96	& 6,13	&8.04	&0.57	&4.56&  80	&0.33	&0.87	&1.30\\
NGC5256		&0.028	&315$^{b}$	&41.85	& 7,13	&6.92	&0.44	&4.29& 214	&0.31	&1.30	&2.23 \\
NGC5347		&0.008	& 93	        &40.45	& 6,13	&6.79	&0.02	&3.67&  90	&0.12	&0.92	&1.31\\
NGC5674		&0.025	&129	        &41.87	&19,16	&7.36	&0.45	&4.49&  68	&0.21	&0.80	&1.87\\
NGC5695		&0.014	&144	        &40.50	& 6,13	&7.56	&0.02	&3.75&  30	&0.18	&0.44	&1.14\\
NGC5929		&0.008	&121	        &40.96	& 6,13	&7.25	&0.06	&4.08&  20	&0.12	&0.27	&2.09\\
NGC7172		&0.009	&154	        &39.77	&19,13	&7.67	&0.01	&3.47& 125	&0.10	&1.06	&2.16\\
NGC7674		&0.029	&144	        &42.49	& 6,13	&7.56	&1.93	&4.72& 120	&0.42	&1.04	&1.88\\
NGC7682		&0.017	&123	       &41.72	& 6,13	&7.28	&0.32	&4.35&  30	&0.24	&0.44	&0.76\\
\hline
\end{tabular}
\end{center}

\vskip 0.1cm
\parbox{5.7in}
{\baselineskip 10pt

\noindent 
$^{a}$the blackhole mass are directly taken from Peterson et al. (2004).\\
$^{b}$refers to [O {\sc iii}] FWHM.\\
$^{c}$based on $\mbh-M_{\rm bulge}$ relation, F01475-0740: $M_{\rm bulge}=-18.80$; NGC 3660: $M_{\rm bulge}=-18.38$.

\vskip 0.3cm
Note.-(1) source name; (2) redshift; (3) FWHM of H$\beta$ for Seyfert 1s or stellar velocity dispersion $\sigma$ for 
Seyfert 2s (in km s$^{-1}$); (4) luminosity of 5100\AA~ deduced from extrapolation of $F_{\nu}\propto \nu^{-0.5}$
or \oiii$\lambda 5007$\AA~ (in erg~s$^{-1}$);
(5) references for columns (3) and (4) are given below, respectively; (6) black hole mass (in $\sunm$); (7) accretion rate 
(in $\sunm$yr$^{-1}$);
(8) gas surface density (in $\sunm$pc$^{-2}$); (9) surface brightness of the 3.3 $\mu$m PAH emission feature
(in $\times10^{39}$ergs s$^{-1}$kpc$^{-2}$); (10) the scale of the starburst regions (in kpc); (11) and (12) are
the lower ($\sigsfr^{\rm L}$) and upper ($\sigsfr^{\rm U}$) limits of surface density of 
star formation rates, respectively (in $\sunm$yr$^{-1}$kpc$^{-2}$).

\vskip 0.3cm
Reference.-(1) NED; (2) Peterson et al. (2004); (3) Spinogilio et al. (1995); (4) Doroshenko \& 
Terebezh (1979); (5) Kinney et al. (1993); (6) Nelson \& Whittle (1995); (7) Dahari \& Robertis (1988); (8) Lipari et al. 
(1991); (9) Corral et al. (2005); (10) Kirhakos \& Steiner (1990); (11) Visvanathan \& Griersmith (1977); 
(12) Cid Fernandes et al. (2004); (13) Gu \& Huang (2002);
(14) Kailey \& Lebofsky (1988); (15) Heraudeau \& Simien (1998); (16) Bassani (1999); (17) Whittle et al. (1988);
(18) Whittle (1992); (19) Garcia-Rissmann et al. (2005); (20) Crenshaw et al. (2003); (21) Marzini et al. (2003); 
(22) Veron-Cetty et al. (2001); (23) Postman \& Lauer (1995).

\vskip 0.3cm

1. NASA/IPAC Extragalatic Database

2. Peterson, B. M., et al. 2004 \apj, 613, 682

3. Spinoglio, L., Malkan, M. A., Rush, B., Carrasco, L. \& Recillas-cruz, E. 1995 \apj, 453, 616

4. Doroshenko, V. T. \& Terebezh, V. Yu. 1979 SvAL, 5, 305

5. Kinney, A. L., Bohlin, R. C., Calzetti, D., Panagia, N., \& Wyse, R. F. G. 1993 \apjs, 86, 5

6. Nelson, C. H. \& Whittle, M. 1995 \apjs, 99, 67

7. Dahari, O. \& Robertis, M. M. D. 1988 \apjs, 67, 249

8. Lipari, S., Bonatto, C. \& Pastoriza, M. 1991 \mnras, 253, 19

9. Corral, A., Barcons, X., Carrera, F. J., Ceballos, M. T., Mateos, S. 2005 \aap, 431, 97

10. Kirhakos, S. D. \& Steiner, J. E. 1990 \aj, 99, 1722

11. Visvanathan, N., Griersmith, D.1977 \aap, 59,317

12. Cid Fernandes, R., et al. 2004 \mnras, 355, 273

13. Gu, Q. \& Huang, J. 2002 \apj, 579, 205

14. Kailey, W. F. \& Lebofsky, M. J. 1988 \apj, 326, 653

15. Heraudeau, P. \& Simien, F. 1998 A\&AS, 133, 317

16. Bassani, L., et al. 1999 \apjs, 121, 473

17. Whittle, M., et al. 1988 \apj, 326, 125

18. Whittle, M. 1992 \apjs, 79, 49

19. Garcia-Rissmann, A., et al. 2005 \mnras, 359, 765

20. Crenshaw, D. M., Kraemer, S. B. \& Gabel, J. R. 2003 \apj, 126, 1690

21. Marzini, P., et al. 2003 \apjs, 145, 199

22. Veron-Cetty, M.-P., Veron, P. \& Gongalves, A.C. 2001 \aap, 372, 730

23. Postman, M., \& Lauer, T. R. 1995  \apj, 440, 28

}

\end{document}